\documentclass[10pt]{article}
\usepackage{amsmath,amsthm,amssymb,amsfonts, fancyhdr, color, comment, graphicx, environ, esint, float, mdframed, stix, subcaption, indentfirst, mathtools}
\usepackage[breakable]{tcolorbox}
\usepackage[top=0.75in, bottom=0.75in, left=0.75in, right=0.75in]{geometry}
\newcommand{\supp}{\mathrm{supp}}

\title{On Phase Retrieval for Continuous and Discrete Fourier Transforms}
\author{Roman G. Novikov and Tianli Xu}
\date{}
\begin{document}
\maketitle
\noindent \textbf{Abstract}: We continue studies on phase retrieval for continuous and discrete Fourier transforms in multidimensions. Using finite difference operators, we give a large class of unexpected examples of non-uniqueness for this problem, including examples with the sparsity condition. A prototype of this construction in the continuous case is given in the work Novikov, Xu (JFAA, 2026), using linear differential operators. The construction of the present work also yields a large class of non-trivial Pauli partners, i.e., different functions with the same intensities in both configuration and Fourier domains. Besides, our construction yields examples that solve an old open question in phase retrieval with background information arising in many areas including Fourier holography. 
\\ \hfill \\
\textbf{Keywords}: continuous and discrete Fourier transforms, phase retrieval, Pauli problem, Fourier holography \\
\textbf{AMS subject classification}: 42A38, 35R30, 39A12, 39A70
\section{Introduction} 
We consider the continuous and discrete Fourier transforms ${F}$ and $\mathcal{F}$ defined as follows:
\begin{equation}
    \hat{u}(p) = Fu(p) = \frac{1}{(2 \pi)^d}\int_{\mathbb{R}^d} e^{ipx}u(x) dx \text{, }  p \in \mathbb{R}^d,
\label{eq:fourier-transform1} 
\end{equation}
\begin{equation}
    \hat{v}(p) = \mathcal{F}v(p) = \frac{1}{(2\pi)^d} \sum_{x \in \mathbb{Z}^d} e^{ipx} v(x), \, p \in \mathbb{T}^d = \mathbb{R}^d/2\pi \mathbb{Z}^d,
    \label{eq:fourier-transform2}
\end{equation}
where $u$, $v$ are complex valued test functions. \\
\indent We consider the following phase retrieval problem in Fourier analysis. 
\\ \hfill \\
\indent \textbf{Problem 1}: \textit{(a) Find $u$ from $|\hat{u}|$ in the continuous case; (b) Find $v$ from $|\hat{v}|$ in the discrete case.}
\\ \hfill \\
\indent Note that the continuous case reduces to the discrete one when 
\begin{equation}
    u(x) = \sum_{y \in \mathbb{Z}^d} v(y) \delta(x - y), \, x \in \mathbb{R}^d,
    \label{function u}
\end{equation}
where $\delta$ is the Dirac delta function. \\
\indent We also consider the following Pauli problem. 
\\ \hfill \\
\indent \textbf{Problem 2}: \textit{(a) Find $u$ from $|u|$ and $|\hat{u}|$ in the continuous case; (b) Find $v$ from $|v|$ and $|\hat{v}|$ in the discrete case.}
\\ \hfill \\
\indent Problems 1 and 2 naturally arises in quantum mechanics, optics, holography, and related areas. Studies on these problems include, in particular, uniqueness, non-uniqueness and reconstruction results. As a relevant literature, see, for example, \cite{BEGM}, \cite{BM}, \cite{CF1} - \cite{Le}, \cite{NS2}, \cite{NX}, \cite{Ro}, \cite{Wa} for Problem 1, and \cite{BT}, \cite{J}, \cite{Pau} for Problem 2, in particular. \\
\indent Usually, Problems 1 and 2 are considered under the assumption that $u$ and $v$ are compactly supported. \\
\indent Recall that the non-uniqueness in Problems 1(a) and 1(b) (respectively), for compactly supported $w = u, v$ (respectively) with possible additional assumptions is considered as non-trivial (and of interest) if it does not reduce to the functions $w_{\alpha, y}$ and $\tilde{w}_{\alpha, y}$ associated to $w$. Here, 
\begin{equation}
   w_{\alpha, y}(x) = e^{i \alpha} w(x - y), 
   \label{Associated function 1}
\end{equation}
\begin{equation}
   \tilde{w}_{\alpha, y}(x) = e^{i \alpha}\overline{w(-x + y)},
   \label{Associated function 2}
\end{equation}
where $\alpha \in \mathbb{R}$, $x, y \in X = \mathbb{R}^d, \mathbb{Z}^d$ (respectively), bar denotes the complex conjugation. In connection with these criteria of non-triviality for non-uniqueness, see, for example, \cite{CF1}, \cite{NX}, at least for Problem 1(a). \\
\indent We also use these criteria for Problem 2. However, the functions $w_{\alpha, y}$ and $\tilde{w}_{\alpha, y}$ in formulas (\ref{Associated function 1}) and (\ref{Associated function 2}) do not give examples of non-uniqueness for Problem 2 in general. In this case, $w_{\alpha, y}$ should be considered only for $y = 0$, and $\tilde{w}_{\alpha, y}$ should be considered only when $|w(-x + y)| = |w(x)|$. \\
\indent In addition, examples of non-uniqueness for Problems 1 and 2 in dimension $d \geq 2$ are considered as non-trivial if they are not reduced to the one-dimensional case via the tensorization
\begin{equation}
    w(x) = \prod_{j = 1}^dw_j(x_j), \, x = (x_1, \cdots, x_d) \in X,
    \label{tensorization}
\end{equation}
in the continuous case, and via a similar tensorization in the discrete case.  \\
\indent In the present work, we give a new very efficient method of constructing non-trivial examples of non-uniqueness for Problems 1 and 2 in dimension $d\geq 1$; see Theorems 1 and 2 in Section 2. \\
\indent Our construction involves finite difference operators, and proceeds from a related construction in \cite{NX}, where linear differential operators with constant coefficients are used. In turn, \cite{NX} continues studies of \cite{BT}, \cite{CF1}, \cite{H}, \cite{J}, \cite{Ro}, \cite{Wa}. In addition, the construction of the present work has some similarity with Example 1 in \cite{J}. \\
\indent In particular, our construction yields examples for the case of compactly supported $v$ and $u$, even under the sparsity condition in multidimensions. \\
\indent In addition, in some sense, Theorems 1 and 2 are the most interesting in the discrete case. The point is that in the compactly supported discrete case, phased or phaseless Fourier data are usually considered on a subset $S$ of $\mathbb{T}^d$,  where the cardinality $\#S$ is finite (see, e.g., \cite{BEGM}, \cite{CSV}), whereas $S = \mathbb{T}^d$ in the counterexamples of Theorems 1 and 2 ! Thus, in this sense, the counterexamples of Theorems 1 and 2 are very strong in the discrete case. \\ 
\indent In the present work, we also consider the following phase retrieval problem with background information formulated below. \\
\indent Let 
\begin{equation}
\begin{array}{c}
D, D_0 \text{ are open, convex, bounded domains in } \mathbb{R}^d, \\
\operatorname{dist}(D, D_0) = R > 0 .
\end{array}
\label{Formulation of problem 3 (1)}
\end{equation}
\indent In the discrete case, we also assume that 
\begin{equation}
    D \cap \mathbb{Z}^d\neq \emptyset, \, D_0 \cap \mathbb{Z}^d \neq \emptyset.
\label{Formulation of problem 3 (2)}
\end{equation}
\indent We assume that
\begin{equation}
\begin{array}{c}
    \supp \, u \subset D, \, \supp \, u_0 \subset D_0, \, u_0 \text{ is a non-zero function on } \mathbb{R}^d\, (\text{in the continuous case}), \\ \supp \, v \subset D \cap \mathbb{Z}^d, \, \supp \, v_0 \subset D_0 \cap \mathbb{Z}^d, \, v_0 \text{ is a non-zero function on } \mathbb{Z}^d \,(\text{in the discrete case}).
\end{array}
\label{Formulation of problem 3 (3)}
\end{equation}
\indent \textbf{Problem 3}: \textit{(a) Find $u$ from $|F(u+u_0)|$ and $u_0$ in the continuous case; (b) Find $v$ from $|\mathcal{F}(v+v_0)|$ and $v_0$ in the discrete case. }
\\ \hfill \\
\indent Problem 3 is a version of Problem 1 with the background information $u_0$ or $v_0$. This problem arises in many areas including Fourier holography and has a long history; see, for example, \cite{BM}, \cite{HNS}, \cite{ML} - \cite{NS2}. \\
\indent In particular, it is known that Problem 3 is uniquely solvable via explicit formulas for the case when $\operatorname{dist}(D, D_0) > \operatorname{diam}(D)$, where 
\begin{equation}
    \operatorname{diam}(D) = \operatorname{sup}_{x,y \in D} |x-y|.
\label{Assumption}
\end{equation}
\indent In the present work, we give examples of non-uniqueness for Problem 3 for the case when  $\operatorname{dist}(D, D_0) < \operatorname{diam}(D)$; see Example 2, Theorems 3 and 4 in Section 2. To our knowledge, this result solves an old open question in phase retrieval with background information. Note also that Example 2 and Theorem 4 are obtained in the framework of the same construction as Theorems 1 and 2. \\
\indent The main results of this work are presented in detail and proved in Sections 3 and 4. 
 \section{Preliminaries}
 \indent In the Fourier domain, the definitions (\ref{Associated function 1}), (\ref{Associated function 2}) of associated functions can be written as follows:
\begin{equation}
    \hat{w}_{\alpha, y}(p) = \hat{w}(p)\operatorname{exp}(i\alpha + iyp) \text{, } p \in P,
\label{Associated function fourier 1}
\end{equation}
\begin{equation}
    \hat{\tilde{w}}_{\alpha, y}(p) = \overline{\hat{w}(p)}\operatorname{exp}(i\alpha + iyp) \text{, } p \in P,
\label{Associated function fourier 2}
\end{equation}
where $P = \mathbb{R}^d$ for the continuous case and $P = \mathbb{T}^d$ for the discrete case. \\
\indent We consider the linear finite difference operators $A$ and $\mathcal{A}$ defined as follows:
\begin{equation}
    (A\psi)(x) = \sum_{y \in T} a_y \psi(x + y), \, x \in \mathbb{R}^d, \indent T \subset \mathbb{R}^d, \, 1 \leq \#T < + \infty, \indent a_y \in \mathbb{C}\setminus \{0\}, 
\label{A1}
\end{equation}
\begin{equation}
    (\mathcal{A}\psi)(x) = \sum_{y \in T} a_y \psi(x + y), \, x \in \mathbb{Z}^d, \indent T \subset \mathbb{Z}^d, \, 1 \leq \#T < + \infty, \indent a_y \in \mathbb{C} \setminus \{0\},
\label{A2}
\end{equation}
where $\psi$ is a complex-valued test function on $\mathbb{R}^d$ in (\ref{A1}) and on $\mathbb{Z}^d$ in (\ref{A2}), $\#$ denotes the cardinality, and $a_y$ are coefficients. \\
\indent Let $A^*$ denote the conjugate operator for $A$ in $L^2(\mathbb{R}^d)$ and $\mathcal{A}^*$ denote the conjugate operator for $\mathcal{A}$ in $L^2(\mathbb{Z}^d)$. Thus,
\begin{equation}
    (A^*\psi)(x) = \sum_{y \in T} \overline{a_y} \psi(x - y), \, x \in \mathbb{R}^d, 
    \label{A*1}
\end{equation}
\begin{equation}
    (\mathcal{A}^*\psi)(x) = \sum_{y \in T} \overline{a_y} \psi(x - y), \, x \in \mathbb{Z}^d, 
    \label{A*2}
\end{equation}
where bar denotes the complex conjugation. \\
\indent Let 
\begin{equation}
    \sigma(x) = (2\pi)^d\sum_{y \in T} a_y \delta(x + y), \, x \in X, \indent \hat{\sigma}(p) = \sum_{y \in T} a_y e^{-ipy}, \, p \in P,
    \label{sigma}
\end{equation}
where $X$ and $P$ are as in (\ref{A1}), (\ref{A2}) and (\ref{A*1}), (\ref{A*2}), and $\delta$ is the Dirac-delta function for $X = \mathbb{R}^d$, whereas $\delta$ is the Kronecker symbol for $X = \mathbb{Z}^d$ (i.e., $\delta$ is the indicator function of $\{0\} \in \mathbb{Z}^d$). \\
\indent Note that 
\begin{equation}
    (FA\psi)(p) = \hat{\sigma}(p) F\psi(p),  \indent (FA^*\psi)(p) = \overline{\hat{\sigma}(p)} F\psi(p), \, p \in \mathbb{R}^d,
    \label{A Fourier 1}
\end{equation}
\begin{equation}
    (\mathcal{F}\mathcal{A}\psi)(p) = \hat{\sigma}(p)\mathcal{F}\psi(p),  \indent (\mathcal{F}\mathcal{A}^*\psi)(p) = \overline{\hat{\sigma}(p)} \mathcal{F}\psi(p), \, p \in \mathbb{T}^d,
    \label{A Fourier 2}
\end{equation}
where $F$,  $\mathcal{F}$ are defined in (\ref{eq:fourier-transform1}), (\ref{eq:fourier-transform2}). 
\section{The main results}
\indent The basic construction of the present work is given in the following theorem. 
\\ \hfill \\
\indent \textbf{Theorem 1}: \textit{Let
\begin{equation}
    f = L\psi, \indent g = L^*\psi \text{ on } X = \mathbb{R}^d \text{ or } X = \mathbb{Z}^d \text{, respectively},
    \label{thm1}
\end{equation}
where $L = A$, $L^* = A^*$ or $L = \mathcal{A}$, $L^* = \mathcal{A}^*$, respectively, as in (\ref{A1}) - (\ref{A*2}), $\psi \in L^2(X)$ is a compactly supported function on $X$. Suppose that 
\begin{equation}
    \text{$\sigma$ defined in (\ref{sigma}) is not associated to itself in the sense of formula (\ref{Associated function 2}) },
    \label{Condition 1 thm 1}
\end{equation}
\begin{equation}
    \text{$\psi$ is not associated to itself in the sense of formula (\ref{Associated function 2}).}
    \label{Condition 2 thm 1}
\end{equation}
\indent Then: \\
\indent (i) $f$ and $g$ are compactly supported functions on $X$,\\
\indent (ii) $|\hat{f}|^2 = |\hat{g}|^2 \nequiv 0 \text{ on } P$, where $P = \mathbb{R}^d \text{ or } \mathbb{T}^d$, respectively, \\
\indent (iii) $f$ and $g$ are not associated functions in the sense of formulas (\ref{Associated function 1}), (\ref{Associated function 2}). }
\\ \hfill \\
\indent Theorem 1 is proved in Section 4.1. \\
\indent Theorem 1 gives a very efficient method of constructing non-trivial examples of non-uniqueness for Problems 1(a) and 1(b) in dimension $d \geq 1$. \\
\indent One of the simplest examples illustrating Theorem 1 is as follows. 
\\ \hfill \\
\indent \textit{Example 1}: \textit{Let 
\begin{equation}
    \sigma(x) = (2\pi)^d \left[a_1 \delta(x) + a_2 \delta(x + y)\right],
    \label{Ex1.1}
\end{equation}
\begin{equation}
    \psi(x) = b_1\chi_r(x) + b_2\chi_r(x - z), 
     \label{Ex1.2}
\end{equation}
\begin{equation}
    \chi_r(x) = 
    \begin{cases}
        1  \text{ if } |x| < r,\\
        0 \text{ if } |x| \geq r,
    \end{cases}
     \label{Ex1.3}
\end{equation}
where $x, y, z \in X$, $r > 0$, $a_1, a_2, b_1, b_2 \in \mathbb{C} \setminus \{ 0 \}$, and $y\neq 0$, $z \neq 0$, $|a_1| \neq |a_2|$, $|b_1| \neq |b_2|$. \\
\indent Then the assumptions of Theorem 1 are fulfilled, and 
\begin{equation}
    f(x) = a_1b_1\chi_r(x) + a_1b_2\chi_r(x - z) + a_2b_1\chi_r(x+ y) + a_2b_2\chi_r(x - z+y),
     \label{Ex1.4}
\end{equation}
\begin{equation}
    g(x) = \overline{a_1}b_1\chi_r(x) + \overline{a_1}b_2\chi_r(x -z) + \overline{a_2}b_1\chi_r(x -y) + \overline{a_2}b_2\chi_r(x - z-y).
     \label{Ex1.5}
\end{equation}
\indent In addition, in order to maximize similarity of supports of $f$ and $g$, it is useful to consider $g_y(x) = g(x+y)$, which is associated to $g$ in the sense of formula (\ref{Associated function 1}): 
\begin{equation}
    g_y(x) = \overline{a_2}b_1\chi_r(x) + \overline{a_2}b_2\chi_r(x - z) + \overline{a_1}b_1\chi_r(x + y) + \overline{a_1}b_2\chi_r(x -z + y).
    \label{1.6}
\end{equation}
\indent In particular, if $z = y$, then 
\begin{equation}
    f(x) = (a_1b_1 + a_2b_2)\chi_r(x) + a_1b_2\chi_r(x - z) + a_2b_1\chi_r(x + z),
    \label{1.7}
\end{equation}
\begin{equation}
    g_y(x) = (\overline{a_1}b_2+\overline{a_2}b_1)\chi_r(x) + \overline{a_2}b_2\chi_r(x - z) + \overline{a_1}b_1\chi_r(x+z).
     \label{Ex1.8}
\end{equation}
\indent In addition, if $b_1 = a_2$, $b_2 = -a_1$, then
\begin{equation}
    f(x) = a_2^2\chi_r(x + z)-a_1^2\chi_r(x - z),
     \label{Ex1.9}
\end{equation}
\begin{equation}
     g_y(x) = \overline{a_1}a_2\chi_r(x+z) + (-|a_1|^2+|a_2|^2)\chi_r(x) -a_1\overline{a_2}\chi_r(x - z).
      \label{Ex1.10}
\end{equation}}
\indent A very important particular case of Theorem 1 consists in the following result. 
\\ \hfill \\
\indent \textbf{Theorem 2}: \textit{Let $f$, $g$, $L$, $\psi$ be as in formula (\ref{thm1}), (\ref{Condition 2 thm 1}).
Suppose that 
\begin{equation}
    T = -T, 
\label{thm2.0}
\end{equation}
\begin{equation}
|a_y| \equiv |a_{-y}|, \indent a_y \not\equiv e^{i\alpha}\overline{a_{-y}}, \, y \in T, \, \alpha \in \mathbb{R}, 
\label{thm2.1}
\end{equation}
\begin{equation}
    (\supp \, \psi - y_i) \cap (\supp \, \psi - y_j) = \emptyset, \,\forall y_i, y_j \in T, i \neq j,
\label{thm2.2}
\end{equation}
where $a_y$ are the coefficients in (\ref{A1}), (\ref{A2}). Then $f$, $g$ satisfy conditions (i) - (iii) of Theorem 1, and 
\begin{equation}
    |f|=|g| \text{ on } X.
\label{thm2.3}
\end{equation}}
\indent \textit{Remark 1: Conditions (\ref{thm2.0}), (\ref{thm2.1}) imply (\ref{Condition 1 thm 1}). } \\ \hfill \\ 
\indent Theorem 2 and Remark 1 are proved in Section 4.2.  \\
\indent Theorem 2 gives a very efficient method of constructing non-trivial examples of non-uniqueness for Problems 2(a) and 2(b) in dimension $d \geq 1$. \\
\indent One of the simplest examples illustrating Theorem 2 is as follows. 
\\ \hfill \\
\indent Let 
\begin{equation}
    \mathcal{B}_\rho(a) = \{ x \in \mathbb{R}^d: \text{ } |x - a| < \rho \}, \text{ } a \in \mathbb{R}^d, \text{ } \rho > 0.
    \label{notation ball}
\end{equation}
\indent \textit{Example 2: Suppose that  
\begin{equation}
    \sigma(x) = (2\pi)^d \left[a_1 \delta(x) + a_2\delta(x + y) + a_{3}\delta(x - y) \right],
    \label{Ex2.1}
\end{equation}
\begin{equation}
    a_1 \in \mathbb{C} \setminus \{0\},\, a_2 \in \mathbb{R} \setminus \{0\}, \indent a_3 = a_2 e^{i\phi},\, a_1 \neq \overline{a_1}e^{i\phi}, \, \phi \in [0, 2\pi),
    \label{Ex2.2}
\end{equation}
$\psi$ is not associated to itself in the sense of formula (\ref{Associated function 2}), and 
\begin{equation}
    \supp \, \psi \subset \mathcal{B}_\rho(a),
    \label{Ex2.3}
\end{equation}
where $a, x,y \in X$, $0 < \rho < |y|/2$. \\
\indent Then the assumptions of Theorem 2 are fulfilled, and 
\begin{equation}
    f(x) = a_1 \psi(x) + a_2 \psi(x+y) + a_2 e^{i\phi}\psi(x-y),
    \label{Ex2.4}
\end{equation}
\begin{equation}
    g(x)e^{i\phi} = \overline{a_1}e^{i\phi}\psi(x) +  a_2 \psi(x+y) + a_2 e^{i\phi}\psi(x-y).
    \label{Ex2.5}
\end{equation}
\indent In particular, here, one can take $\psi$ as in Example 1, assuming that 
\begin{equation}
    z = 2a,\indent |z| < 2\rho, \indent  r \leq \rho - \frac{|z|}{2}.
    \label{Ex2.6}
\end{equation}
}
\indent Next, we construct different examples of non-uniqueness for Problems 3(a) and 3(b), i.e., we construct functions $w_0$, $w_1$, $w_2$ on $X$ such that 
\begin{equation}
    \supp \, w_0 \subset D_0, \indent \supp \, w_j \subset D, \, j = 1, 2, \indent \operatorname{dist}(D, D_0) = R, \indent 0< R < \operatorname{diam}(D),
    \label{Pb3.1}
\end{equation}
\begin{equation}
    |\widehat{w_1 + w_0}|^2 \equiv |\widehat{w_2 + w_0}|^2 \text{ on } P,
    \label{Pb3.2}
\end{equation}
\begin{equation}
    w_0 \neq 0, \, w_1 \neq w_2 \text{ in } L^2(X), 
    \label{Pb3.3}
\end{equation}
where $D$, $D_0$ are as in (\ref{Formulation of problem 3 (1)}), (\ref{Formulation of problem 3 (2)}), $X = \mathbb{R}^d$, $P = \mathbb{R}^d$ in the continuous case and $X = \mathbb{Z}^d$, $P = \mathbb{T}^d$ in the discrete case. In addition, in an important part of our examples, we also have that 
\begin{equation}
    w_1 + w_0 \text{ and } w_2 + w_0 \text{ are not associated functions in the sense of formulas (\ref{Associated function 1}), (\ref{Associated function 2})}.  
    \label{re2}
\end{equation}
\indent A very interesting point of Example 2 consists in the observation that formulas (\ref{Ex2.4}), (\ref{Ex2.5}) give also an example of non-uniqueness for Problems 3(a) and 3(b) for the case when $\operatorname{dist}(D, D_0) < \operatorname{diam}(D)$ ! One can see this using the following notations:
\begin{equation}
    w_0(x) = a_2e^{i\phi}\psi(x-y),
    \label{w0}
\end{equation}
\begin{equation}
    w_1(x) = a_1 \psi(x) + a_2 \psi(x+y), \indent w_2(x) = \overline{a_1}e^{i\phi}\psi(x) +  a_2 \psi(x+y),
    \label{w1}
\end{equation}
\begin{equation}
    {D_0} = \mathcal{B}_{\rho}\left(a + y\right), \indent {D} = \operatorname{ch}(\mathcal{B}_\rho(a) \cup \mathcal{B}_\rho(a-y)),
    \label{w2}
\end{equation}
where $\operatorname{ch}(A)$ is the convex hull of a set $A$ in $\mathbb{R}^d$. The point is that in this setting, formulas (\ref{Pb3.1}) - (\ref{re2}) hold and, in addition, 
\begin{equation}
    R = |y| - 2\rho >0,
    \label{re1}
\end{equation}
\indent To our knowledge, examples of non-uniqueness for Problems 3(a) and 3(b) are not available in the literature. In addition, such examples seem to be of great interest even without assumptions that $w_1 + w_0$ and $w_2 + w_0$ are not associated functions in the sense of formulas (\ref{Associated function 1}), (\ref{Associated function 2}). \\
\indent In particular, if, in Example 2, $\psi$ is as in (\ref{Ex1.2}), (\ref{Ex1.3}) with $b_1 \in \mathbb{C}\setminus\{0\}$, $b_2 = 0$, $|z| = 0$, then $w_1 + w_0$ and $w_2 + w_0$ are associated functions in the sense of formula (\ref{Associated function 2}), but formulas (\ref{Pb3.1}) - (\ref{Pb3.3}) and (\ref{w0}) - (\ref{re1}) hold. In fact, this simple example is a particular example of Theorem 3 given below. 
\\ \hfill \\
\indent Let $U_0$, $U_1$ be open bounded convex domains in $\mathbb{R}^d$ such that 
\begin{equation}
    \indent  U_1 = -U_1, \indent \operatorname{dist}(U_1, U_0) = R > 0.
    \label{notation for thm3 1}
\end{equation}
\indent Let 
\begin{equation}
    \tilde{w}(x) = \overline{w(-x)},
    \label{notation for thm3 2}
\end{equation}
where $w$ is a test function on $X$.
\\ \hfill \\
\indent \textbf{Theorem 3}: \textit{Let $\psi$ and $\phi$ be non-zero functions in $L^2(X)$ such that
\begin{equation}
    \supp \, \psi \subset U_0, \, \supp \, \phi \subset U_1, \indent \phi \neq \tilde{\phi}.
    \label{thm3.1}
\end{equation}
\indent Let 
\begin{equation}
    D_0 = U_0, \indent D = \operatorname{ch}\left(U_1 \cup (-U_0)\right),
    \label{thm3.2}
\end{equation}
\begin{equation}
    w_0 = \psi, \indent w_1 = \tilde{\psi} + \phi, \indent w_2 = \tilde{\psi} + \tilde{\phi}.
    \label{thm3.3}
\end{equation}
\indent Then formulas (\ref{Pb3.1}) - (\ref{Pb3.3}) hold, although $w_1 + w_0$ and $w_2 + w_0$ are associated functions in the sense of formula (\ref{Associated function 2}).}
\\ \hfill \\
\indent Theorem 3 is proved in Section 4.3. This proof is elementary. However, we found the formulation of Theorem 3 proceeding from Example 2 illustrating Theorem 2. \\
\indent In addition, Example 2 is strongly generalized by the following result. 
\\ \hfill \\
\indent \textbf{Theorem 4}: \textit{Let $f$, $g$, $L$, $\psi$ be as in formula (\ref{thm1}), where $\psi$ is a non-zero function, and 
\begin{equation}
    \supp \, \psi \subset \mathcal{B},
    \label{thm4.1}
\end{equation}
where $\mathcal{B}$ is an open convex bounded domain in $\mathbb{R}^d$. Suppose that there is $y^* \in T$ such that $y^{*} \in (-T)$ and 
\begin{equation}
    D_0 \cap D = \emptyset, \indent \text{where } D_0 = \mathcal{B} + y^*, \, D = \operatorname{ch}\left(\bigcup_{y \in \left(T \cup (-T)\right)\setminus \{y^*\}} \left(\mathcal{B} + y \right)\right).
    \label{4.5}
\end{equation}
\indent We also assume that 
\begin{equation}
   a_{-y^*} = e^{i\phi}\overline{a_{y^*}}, \, \phi \in [0, 2\pi),
    \label{4.3}
\end{equation}
and that conditions (\ref{Condition 1 thm 1}), (\ref{Condition 2 thm 1}) are fulfilled. Let 
\begin{equation}
    w_0(x) = a_{-y^*}\psi(x-y^*), \indent w_1(x) = f(x) - w_0(x), \indent w_2(x) = g(x)e^{i\phi} - w_0(x).
    \label{4.6}
\end{equation}
\indent Then formulas (\ref{Pb3.1}) - (\ref{re2}) hold.
} \\ \hfill \\
\indent \textit{Remark 2: In Theorem 4, without assumption (\ref{Condition 1 thm 1}), we still have that properties (\ref{Pb3.1}) - (\ref{Pb3.3}) hold. }
\\ \hfill \\
\indent Theorem 4 and Remark 2 are proved in Sections 4.4 and 4.5.  
\\ \hfill \\
\indent \textit{Remark 3: Suppose that assumptions (\ref{thm2.0}), (\ref{4.3}) hold, where $y^* \in T$, and, in addition, 
\begin{equation}
    a_{-y} \not\equiv e^{i\phi}\overline{a_y}, \, y \in T.
    \label{4.4}
\end{equation}
\indent Then condition (\ref{Condition 1 thm 1}) holds.}
\\ \hfill \\
\indent The proof of Remark 3 is completely similar to the proof of Remark 1. \\
\indent In connection with Theorems 1 and 4, the following remark is also of interest. 
\\ \hfill \\
\indent \textit{Remark 4: Let $T$ and $a_y$ be as in (\ref{A1}), (\ref{A2}), (\ref{sigma}), and
$T \neq -T + z$, where $z \in \mathbb{R}^d$. Then condition (\ref{Condition 1 thm 1}) holds.}
\\ \hfill \\
\indent Remark 4 follows from formula (\ref{proof2.2}), which is used in the proof of Theorem 2. \\
\indent In connection with Theorem 4, the following remark is also of interest. 
\\ \hfill \\
\indent \textit{Remark 5: Let $y^*$ satisfy the assumptions formulated in Theorem 4, and $T \neq -T$. Then $T \neq -T+z,\, z \in \mathbb{R}^d$.} \\
\indent Remark 5 is proved in Section 4. 
\\ \hfill \\
\indent \textit{Remark 6: Examples of non-uniqueness for Problem 3 for $d = 1$ can be also obtained proceeding from Example 1 in \cite{J}, although Problem 3 is not considered in \cite{J}. }
\\ \hfill \\
\indent \textit{Remark 7: Constructions of functions $f$ and $g$ in Theorems 1, 2, 4 can be considered in term of zero-flipping between $\hat{f}$ and $\hat{g}$ (in terminology going back to \cite{Wa}), in view of formulas (\ref{A Fourier 1}), (\ref{A Fourier 2}). An interesting point is that already, in Examples 1 and 2, for $d = 1$, this zero-flipping involves infinitely many zeros. This issue will be studied in more detail elsewhere. }
\\ \hfill \\
\indent \textit{Remark 8: In the continuous case, constructions of functions $f$ and $g$ in Theorems 1 and 4 can be further generalized, taking into account Theorem 3.3 of \cite{NX}. We can consider $L$ in formula (\ref{thm1}) in the following form:
\begin{equation}
    L\psi(x) = \sum_{y\in T} a_y (P_y\psi)(x + y),\indent L^*\psi(x) = \sum_{y\in T} \overline{a_y} (P_y^*\psi)(x - y),\indent 1 \leq \#T < + \infty, \indent a_y \in \mathbb{C}\setminus \{0\}, 
\end{equation}
where each $P_y$ is a linear differential operator in $x$ with constant coefficients, $P_y^*$ is adjoint to $P_y$, and $\psi$ is assumed to be smooth and of compact support. This issue will be studied in more detail elsewhere.
}
\section{Proofs of Theorems 1 - 4 and Remarks 2, 5}
\subsection{Proof of Theorem 1}
\indent Item (i) follows from  (\ref{A1}) - (\ref{A*2}). \\
\indent Next, according to our assumptions on $\sigma$ and $\psi$, we have that 
\begin{equation}
     \hat{\sigma}(p) \not\equiv \overline{\hat{\sigma}(p)}\operatorname{exp}(i\alpha + iyp) \text{, } p \in P,
     \label{proof1.1}
\end{equation}
\begin{equation}
    \hat{\psi}(p) \not\equiv \overline{\hat{\psi}(p)}\operatorname{exp}(i\alpha + iyp) \text{, } p \in P,
    \label{proof1.2}
\end{equation}
\begin{equation}
    \hat{\sigma} \nequiv 0 \text{ on } P, \indent \hat{\psi} \nequiv 0 \text{ on } P.
    \label{proof1.3}
\end{equation}
\indent Recall also that
\begin{equation}
    \operatorname{mes} \left( \{ p \in P \text{ }: \text{ } \hat{w}(p) = 0\}\right) = 0,
    \label{proof1.4}
\end{equation}
 where $\hat{w}$ is a non-zero real analytic function on $P$, for example, the Fourier transform of a non-zero compactly supported function $w$ on $X$, and $\operatorname{mes}$ denotes the Lebesgue measure in $P$. \\
 \indent Item (ii) follows from (\ref{A Fourier 1}), (\ref{A Fourier 2}), (\ref{proof1.3}), (\ref{proof1.4}). \\
 \indent The proof of item (iii) is as follows. \\
\indent In view of (\ref{Associated function fourier 1}), (\ref{Associated function fourier 2}), (\ref{A Fourier 1}), (\ref{A Fourier 2}), we have that $f$ and $g$ are associated functions in the sense of formula (\ref{Associated function 1}) if and only if 
\begin{equation}
    \overline{{\hat{\sigma}}({p})}\hat{\psi}(p) = {\hat{\sigma}}(p) \hat{\psi}(p)\operatorname{exp}(i\alpha + iyp) \text{, } p \in P,
    \label{proof1.5}
\end{equation}
and in the sense of formula (\ref{Associated function 2}) if and only if
\begin{equation}
    \overline{{\hat{\sigma}}({p})}\hat{\psi}(p) = \overline{{\hat{\sigma}}\left({p}\right) \hat{\psi}({p})}\operatorname{exp}(i\alpha + iyp) \text{, } p \in P,
    \label{proof1.6}
\end{equation}
for some $\alpha \in \mathbb{R}$ and $y \in X$. \\
\indent One can see that equality (\ref{proof1.5}) is impossible, using (\ref{proof1.1}), (\ref{proof1.4}). \\
\indent One can see that equality (\ref{proof1.6}) is impossible, using (\ref{proof1.2}), (\ref{proof1.4}). \\
\indent Theorem 1 is proved.
\subsection{Proof of Theorem 2}
As mentioned in Remark 1, we have that 
\begin{equation}
    \sigma \not \equiv \tilde{\sigma}_{\alpha,y}, 
    \label{proof2.1}
\end{equation}
where $\tilde{\sigma}_{\alpha,y}$ is defined according to (\ref{Associated function 2}) and $\sigma$ is defined in (\ref{sigma}). \\
\indent Formula (\ref{proof2.1}) follows from properties (\ref{thm2.0}), (\ref{thm2.1}), the properties that 
\begin{equation}
    \supp \, \sigma = -T, \indent \supp \, \tilde{\sigma}_{\alpha,y} = T + y.
    \label{proof2.2}
\end{equation}
and the observation that the equality $T = T + y$ is impossible for $y \neq 0$. Thus, Remark 1 is proved. \\
\indent Our assumptions in Theorem 2, property (\ref{proof2.1}), and Theorem 1 imply that statements (i) - (iii) of Theorem 1 hold. \\
\indent Next, due to (\ref{A*1}), (\ref{A*2}), (\ref{thm2.0}), we have that 
\begin{equation}
    (A^*\psi)(x) = \sum_{y \in T} \overline{a_{-y}} \psi(x + y), \, x \in \mathbb{R}^d, 
    \label{proof2.3}
\end{equation}
\begin{equation}
    (\mathcal{A}^*\psi)(x) = \sum_{y \in T} \overline{a_{-y}} \psi(x + y), \, x \in \mathbb{Z}^d.
    \label{proof2.4}
\end{equation}
\indent Formula (\ref{thm2.3}) follows from (\ref{A1}), (\ref{A2}), (\ref{proof2.3}), (\ref{proof2.4}), the assumption that $|a_y| \equiv |a_{-y}|$, and the condition (\ref{thm2.2}).\\
\indent Theorem 2 is proved. 
\subsection{Proof of Theorem 3}
\indent Formula (\ref{Pb3.1}) follows from (\ref{notation for thm3 1}) - (\ref{thm3.3}).\\
\indent In view of (\ref{thm3.3}), we have that  
\begin{equation}
    w_2(x) + w_0(x) = \overline{w_1(-x) + w_0(-x)}, \, x \in X.
\label{proof3.1}
\end{equation}
\indent Thus, $w_2 + w_0$ is associated to $w_1 + w_0$ in the sense of formula (\ref{Associated function 2}) with $\alpha  = 0$, $y= 0$. This implies formula (\ref{Pb3.2}).\\
\indent The property that $w_0 \neq 0$ follows from its definition in (\ref{thm3.3}) and the assumption $\psi \neq 0$. The property that $w_1 \neq w_2$ follows from their definitions in (\ref{thm3.3}) and the assumption that $\phi \neq \tilde{\phi}$ in (\ref{thm3.1}). Thus, properties (\ref{Pb3.3}) are proved. \\
\indent Theorem 3 is proved. 
\subsection{Proof of Theorem 4}
\indent In view of formulas (\ref{A1}) - (\ref{A*2}), (\ref{thm1}), (\ref{thm4.1}) and the assumptions that $y^* \in T$, $y^* \in (-T)$, we have that 
\begin{equation}
    \supp \, f \subset \bigcup_{y \in T } \left(\mathcal{B} - y \right) = (\mathcal{B} + y^*) \cup \left(\bigcup_{y \in T \setminus \{-y^*\} }  \left(\mathcal{B} - y \right)\right)=(\mathcal{B} + y^*) \cup \left(\bigcup_{y \in(- T) \setminus \{y^*\} }  \left(\mathcal{B} + y \right)\right), 
\label{Proof 4.1}
\end{equation}
\begin{equation}
     \supp \, g \subset \bigcup_{y \in (-T) } \left(\mathcal{B} - y \right) = (\mathcal{B} + y^*) \cup \left(\bigcup_{y \in (-T) \setminus \{-y^*\} }  \left(\mathcal{B} - y \right)\right) = (\mathcal{B} + y^*) \cup \left(\bigcup_{y \in T \setminus \{y^*\} }  \left(\mathcal{B} + y \right)\right).
     \label{Proof4.2}
\end{equation}
\indent In addition, using also (\ref{4.5}), we have that 
\begin{equation}
    w_0 = \chi_{D_0}f = \chi_{D_0}e^{i\phi}g, \indent w_1 = \chi_{D}f, \indent w_2 = \chi_{D}e^{i\phi}g,
    \label{Proof 4.3}
\end{equation}
where $\chi_{D_0}$ and $\chi_D$ are the characteristic functions on the domains $D_0$ and $D$. \\
\indent Properties (\ref{Pb3.1}) follows from (\ref{4.5}), (\ref{Proof 4.1}) - (\ref{Proof 4.3}). \\
\indent In addition, properties (\ref{Pb3.2}) - (\ref{re2}) follow from Theorem 1. \\
\indent Theorem 4 is proved. 

\subsection{Proof of Remark 2}
\indent Remark 2 follows from the proofs of Theorem 4 and Theorem 1, using, in particular, formulas (\ref{4.6}) and the observation that $f$ and $g$ are not associated functions in the sense of formulas (\ref{Associated function 1}), (\ref{proof1.5}), in view of formulas  (\ref{proof1.1}), (\ref{proof1.4}). 

\subsection{Proof of Remark 5}
\indent Suppose that $T = -T + z$, $z \in \mathbb{R}^d \setminus \{ 0\}$. Let
\begin{equation}
    y_1 = -y^* + z, \indent y_2 = y^* + z, \indent y_3 = -y_1 = y^* - z, \indent y_4 = -y_2 = -y^* -z.
    \label{proof6.1}
\end{equation}
\indent By our assumptions, we have that
\begin{equation}
    \{y_1,\, y_2, y_3,y_4\} \subset T \cup \left(-T\right), \indent   y^* \in \operatorname{ch}\left(\{y_2, y_3\}\right) = [y_2, y_3], \indent y_2 \neq y^*, \indent y_3 \neq y^*.
    \label{proof6.2}
\end{equation}
\indent On the other hand, as a corollary of assumption (\ref{4.5}), we have that
\begin{equation}
    y^* \not \in \operatorname{ch} \left(T \cup \left(-T\right)\setminus\{y^*\}\right).
    \label{proof6.3}
\end{equation}
\indent However, formulas (\ref{proof6.2}) and (\ref{proof6.3}) are in contradiction. \\
\indent Remark 5 is proved.

Roman G. Novikov, CMAP, CNRS, École polytechnique, \\
Institut Polytechnique de Paris, 91120 Palaiseau, France\\
\& IEPT RAS, 117997 Moscow, Russia \\
E-mail: novikov@cmap.polytechnique.fr
\\ \hfill \\
Tianli Xu, École polytechnique, \\
Institut Polytechnique de Paris, 91120 Palaiseau, France\\
E-mail: tianli.xu@polytechnique.edu 
 
\end{document}